\documentclass[letterpaper]{IEEEtran}
\usepackage{graphicx,times,amsmath,balance,amsfonts,cite}
\title{The Worst Case ISI channels and the Uniqueness of the Corresponding Minimum Eigenvalue}
\author{Nandana Rajatheva, \IEEEmembership{Senior Member,~IEEE}
\thanks{Author is with the Department of Communication Engineering, University of Oulu, Finland (e-mail: rrajathe@ee.oulu.fi).}}
\begin{document}
\maketitle
%
%
\begin{abstract}
Intersymbol interference (ISI) is a major cause of degradation in the receiver performance of high-speed data communications systems. This arises mainly due to limited bandwidth available. The minimum Euclidean distance between any two symbol sequences is an important parameter in this case at moderate to high signal to noise ratios. It is proven here that as ISI increases the minimum distance strictly decreases when the worst case scenario is considered. From this it follows that the minimum eigenvalue of the worst case ISI channel of a given length is unique.

%
\end{abstract}
%
%
%
\section{Introduction}

Reliable communication among different systems and persons is an essential requirement given the urgent need for information. As the amount of information to be exchanged increases the required data rates also increases. Increased data rates put considerable strains on the resources of the communication systems. One of the scarce resources is the available bandwidth. The bandwidth is always available in limited quantities. Therefore, high-speed communication in a limited bandwidth environment is an important subject for research. It is well known that under these conditions intersymbol interference (ISI) is a major cause of performance degradation. In the worst ISI case, this can be formulated as a symmetric eigenvalue problem for positive definite matrices.
\par
A finite length ISI channel can be represented as a discrete time transversal filter for a given channel impulse response \cite{forney}. An important parameter that defines the receiver performance in terms of bit error rate (BER) versus signal to noise ratio (SNR), is the minimum Euclidean distance ($d_{min}^2$) between any two symbol sequences. It can be shown that certain channel responses give the worst $d_{min}^2$ \cite[Ch. 10,pp. 593-601]{proakis}. The variation of $d_{min}^2$  against the length of ISI was only considered for a few cases without any generalizations.
In this letter it is shown that $(d_{min}^2)_{worst}$ strictly decreases with ISI. Hence it is clear that when the ISI is large the deterioration of the receiver could be very high with a limiting channel capacity of zero for certain channel responses. 

This letter is organized as follows. Section \ref{SysModel} summarizes the system model of interest. Section \ref{WrstDist} consider the specific distance properties of channels. 
The concluding remarks are given in Section \ref{Conclusion}.
%
\section{System Model}
\label{SysModel}
The communication system under consideration,in terms of the received signal-discrete output, is as follows.

\begin{equation}
\label{rxsig}
z_k= \displaystyle\sum\limits_{i=0}^{L-1}f_ia_{k-i}+n_k
\end{equation}

where the noise $n_k$ is white and the channel response is given by $\mathbf{f}^T=[f_0 f_1.....f_{L-1}]$.
\par
The receiver uses the Viterbi algorithm to carry out maximum likelihood sequence estimation ($MLSE$). The distance between two symbol sequences can be given by \cite[Ch. 10,pp. 593-601]{proakis}
\begin{equation}
\label{distance}
d^2= \mathbf{f}^T\mathbf{A}\mathbf{f}
\end{equation}
It is obtained based on the fact, that the distance can be represented using the difference (error) symbols mapped through the ISI trellis diagram. An error event of length $l$ is considered. The matrix $\mathbf{A}$ is the error correlation matrix. The elements are thus obtained from error symbol sequence \cite[Ch. 10,pp. 593-601]{proakis}. $\mathbf{A}$ is a toeplitz matrix with the main diagonal element being the largest. It is also assumed that the channel energy is normalized, i.e., $\mathbf{f}^T\mathbf{f} = 1$. In the following we discuss the properties of worst case ISI channel for a given length of interference as this would give us an idea about the maximum performance loss.

\section{Worst Minimum Distance}
\label{WrstDist}

To obtain the worst minimum distance one simply looks at the eigenvalues of matrix $\mathbf{A}$ for all possible error symbol sequences.  It was shown that a worst case ISI channel is obtained when the eigenvalue of $\mathbf{A}$ is the minimum. The channel response is given by the corresponding eigenvector.
\begin{equation}
	 (d_{min}^2)_{worst}=\mathbf{f}^T_w\lambda_{min}\mathbf{f}_w
                        =\lambda_{min}\mathbf{f}^T_w\mathbf{f}_w
                        =\lambda_{min}
\end{equation}

since $\mathbf{f}^T_w\mathbf{f}_w= 1$.
Recently the author in \cite{sun} considered the bounds of the minimum eigenvalue of positive definite matrices. Here it is considered in terms of the Euclidean distance between symbol sequences of an ISI channel. The question is how does this minimum eigenvalue, or in other words, the worst Euclidean distance, vary with the increase of ISI length. It can be easily shown that it is either equal to or less than the value for a smaller ISI length. However, it is our intention to show that this minimum eigenvalue monotonically (strictly) decreases with ISI.
\par
\par
$\textbf{Theorem}$: The worst $d_{min}^2$ or $(d_{min}^2)_{worst}$ strictly decreases with the length of ISI.
\par

Proof: We prove the above theorem using two separate results. First we show that
\begin{equation}
	 [(d_{min}^2)_{worst}]_{L+1} \leq [(d_{min}^2)_{worst}]_L
\end{equation}

\par
Consider a channel of length $L+1$ with the worst distance given by a certain error event. The error correlation matrix $\mathbf{A}_{(L+1)w}$ corresponding to that error event will have $L+1$ eigenvalues. Now consider a channel of length L and an error event identical to the one mentioned earlier. It is clear that $\mathbf{A}_{Lw}$ is the same as the principal submatrix of $\mathbf{A}_{(L+1)w}$.
>From matrix theory for positive definite matrices, the eigenvalues of the principal submatrix of $\mathbf{A}_{(L+1)w}(=\mathbf{A}_{Lw} )$ separate the eigenvalues of $\mathbf{A}_{(L+1)w}$. Thus considering the minimum eigenvalues of both matrices the first result is obtained.

Now we have to show that
\begin{equation}
	 [(d_{min}^2)_{worst}]_{L+1} < [(d_{min}^2)_{worst}]_L
\end{equation}
	
\par

Consider the channel which gives the worst distance for length $L$. The channel coefficients come from the eigenvector corresponding to the minimum eigenvalue ($ \lambda_{minL}$) of $\mathbf{A}_{Lw}$.
Now consider a channel of length $L+1, \mathbf{f}^T_{(L+1)a}=[f_0 f_1.....f_{L-1} f_L] $ as an augmented channel with the first $L$ coefficients $\mathbf{f}^T_{Lw} =[f_0 f_1.....f_{L-1}] $ being equal to the eigenvector for $A_{Lw}$ corresponding to  $\lambda_{minL}$. The last coefficient $f_L$ is a variable with magnitude greater than zero. This, however, violates the energy condition as $\mathbf{f}^T_{(L+1)a}\mathbf{f}_{(L+1)a} > 1$ for $L+1$. If the same error event $\epsilon_l$ is considered with an additional zero error symbol it can be considered an an error event for ISI length $L+1$. We can also obtain the matrix $\mathbf{A}_{(L+1)a}$. It  is clear from the earlier discussion that $\mathbf{A}_{Lw}$ is the principal submatrix of $\mathbf{A}_{(L+1)a}$. The difference will be in the $(L+1)$th row and column. If matrix $\mathbf{A}_{Lw}$ is given by

\begin{equation}
\label{alw}
	\mathbf{A}_{Lw} = \left[ \begin{array}{ccccc}
\beta_0 & \beta_1 & . & . & \beta_{L-1} \\
\beta_1 & \beta_0 & \beta_1 & . & . \\
\beta_2 & \beta_1 & \beta_0 & . & . \\
. & . & . & . & . \\
\beta_{L-1} & . & \beta_2 & \beta_1 & \beta_0 \end{array} \right]
\end{equation}

The matrix $\mathbf{A}_{(L+1)a}$ is given by \eqref{ala} and  $L$ being a new element which did not appear in $\mathbf{A}_{Lw}$.
\begin{equation}
\label{ala}
	\mathbf{A}_{(L+1)a} = \left[ \begin{array}{cccccc}
\beta_0 & \beta_1 & \beta_2 & . & \beta_{L-1} & \beta_L \\
\beta_1 & \beta_0 & \beta_1 & . & . & \beta_{L-1} \\
\beta_2 & \beta_1 & \beta_0 & . & . & . \\
. & . & . & . & . & \beta_2 \\
\beta_{L-1} & . & . & . & . & \beta_1 \\
\beta_L & \beta_{L-1} & . & \beta_2 & \beta_1 & \beta_0 \end{array} \right] 	
\end{equation}

Now consider
\begin{equation}
\label{dla}
\begin{split}
	d^2_{(L+1)a} &= \mathbf{f}^T_{(L+1)a} \mathbf{A}_{(L+1)a}\mathbf{f}_{(L+1)a} \\	
	&= f^T_{Lw} \mathbf{A}_{Lw}f_{Lw}+f_L(\beta_0f_L+2\beta_Lf_0+2 \displaystyle\sum\limits_{i=1}^{L-1}\beta_{L-i}f_i) \\
  	&= [(d^2_{min})_{worst}]_L +  f_L(\beta_0f_L+2\beta_Lf_0+2 \displaystyle\sum\limits_{i=1}^{L-1}\beta_{L-i}f_i)	
\end{split}
\end{equation}

 Claim: The term  $\beta_Lf_0+ \displaystyle\sum\limits_{i=1}^{L-1}\beta_{L-i}f_i \neq 0$.

$\mathbf{f}^T_{Lw}= [f_0 f_1.....f_{L-1}]$  is an eigenvector of $\mathbf{A}_{Lw}$ as the solution for the $L$ linear relations associated with $\lambda_{minL}$ for that matrix. The above term then presents another linear relation for $[f_0 f_1.....f_{L-1}]$. We can formulate this in the following way. If that term is zero, then we can easily see that $\mathbf{f}^T_{Lw}$ should also be an eigenvector of the matrix with the eigenvalue $\lambda_{minL}$,
\begin{equation}
	\mathbf{A}_L = \left[ \begin{array}{ccccc}
\beta_0 & \beta_1 & . & . & \beta_{L-1} \\
\beta_1 & \beta_0 & \beta_1 & . & . \\
\beta_2 & \beta_1 & \beta_0 & . & . \\
. & . & . & . & . \\
\beta_L & . & \beta_3 & \beta_1 & \beta^1_1 \end{array} \right]
\end{equation}

This is obtained by replacing the last row of $\mathbf{A}_{Lw}$ by the  $\beta's$ given in \eqref{alw} with $\beta^1_1=\beta_1+\lambda_{minL}$. It is different to $\mathbf{A}_{Lw}$ in the last row, thus making them different matrices. Therefore, the eigenvalues need to be different with different eigenvectors as we are solving two sets of equations. The matrices are not related by any similarity transformation which produce the same eigenvalues. The eigenvectors of such matrices are still different.
\par
If, on the other hand, we argue that both these rows are the same, producing the same eigenvalues and vectors then the following should be satisfied;

\begin{equation}
\begin{split}
\beta_L&=\beta_{L-1}=\beta_{L-2}=....\beta_2=\beta_1\\
\beta_0&=\beta_1+\lambda_{minL}
\end{split}
\end{equation}

We already have the error correlation matrix for $L$ and for the known cases of the worst case scenario, this condition is not satisfied. In most cases $\beta_0$ needs to be an integer. Thus using an inductive method we can eliminate this possibility for all cases.

\par
Thus, we have arrived at a contradiction.

Hence  $\beta_Lf_0+\displaystyle\sum\limits_{i=1}^{L-1}\beta_{L-i}f_i \neq 0$.

This implies that $d_{(L+1)a}^2$ can be made smaller than $[(d_{min}^2)_{worst}]_L$  by judicious selection of $f_L$ in \eqref{dla}. Therefore even with $\mathbf{f}^T\mathbf{f} > 1$ for $L+1$,

\begin{equation}
	  d_{(L+1)a}^2 < [(d_{min}^2)_{worst}]_L.
\end{equation}

Thus with the channel coefficients normalized so that $\mathbf{f}^T\mathbf{f} = 1$ for $L+1$, it is clear

\begin{equation}
	  d_{(L+1)a}^2 < [(d_{min}^2)_{worst}]_L.
\end{equation}

This completes the second part of the proof. It is immediately implied that,

\begin{equation}
[(d_{min}^2)_{worst}]_{L+1} \leq (d_{min}^2)_{(L+1)a}	  	
\end{equation}

and hence	

\begin{equation}
[(d_{min}^2)_{worst}]_{L+1} < (d_{min}^2)_L	  	
\end{equation}

\textbf{QED}.
\par
$\textbf{Corollary I}$: The minimum eigenvalue of the correlation matrix corresponding to the error event of the worst ISI channel is unique..
\par
Proof: It follows from the theorem that the worst distance strictly decreases with ISI. Since worst distance always corresponds to the minimum eigenvalue of a certain error event for that length of ISI, it is unique for that $L$.
\par
$\textbf{Corollary II}$: The roots of the worst ISI channel lie on the unit circle.
\par
Proof: It follows from the uniqueness of the eigenvalue and shown in \cite{larsson}, if Corollary I holds, which we showed as above.

Although we considered only real valued coefficients and matrices the same applies to channels with complex coefficients and transmitted (or error) symbols \cite{larsson}.

%

\section{Conclusion}
\label{Conclusion}
We presented a theorem which established the fact that the worst Euclidean distance strictly decreases with the length of ISI. This in turn proves that the minimum eigenvalue of the worst case ISI channel of any given length is unique.

\end{document}